\documentclass[ reprint,amsmath,amssymb,aps,twocolumn,superscriptaddress]{revtex4-1}

\usepackage[nopar]{lipsum}
\usepackage{graphicx}% Include figure files
\usepackage{dcolumn}% Align table columns on decimal point
\usepackage{bm}% bold math
\usepackage[capitalize]{cleveref}
\let\revappendix\appendix

\usepackage{siunitx}

\begin{document}

\preprint{APS/123-QED}

\title{High mobility transport in isotopically-enriched ${^{12}}$C and ${^{13}}$C exfoliated graphene}
%\date{\today}
%\usepackage{color}
%\usepackage{graphicx,textcomp,amssymb,amsmath,dcolumn,hyperref}
%\usepackage{siunitx}
%\usepackage{comment}

\author{Shuichi Iwakiri}
\email{siwakiri@phys.ethz.ch}
\author{Jakob Miller}
\author{Florian Lang}
\author{Jakob Prettenthaler}
\affiliation{Solid State Physics Laboratory, ETH Zurich,~CH-8093~Zurich, Switzerland}
\author{Takashi Taniguchi}
\affiliation{Research Center for Materials Nanoarchitectonics, National Institute for Materials Science,  1-1 Namiki, Tsukuba 305-0044, Japan}
\author{Kenji Watanabe}
\affiliation{Research Center for Electronic and Optical Materials, National Institute for Materials Science, 1-1 Namiki, Tsukuba 305-0044, Japan}
\author{Sung Sik Lee}
\affiliation{ScopeM, ETH Zurich,~CH-8093~Zurich, Switzerland}
\author{Pascal Becker}
\author{Detlef G{\"u}nther}
\affiliation{Department of Chemistry and Applied Biosciences, ETH Zurich,~CH-8093~Zurich, Switzerland}
\author{Thomas Ihn}
\affiliation{Solid State Physics Laboratory, ETH Zurich,~CH-8093~Zurich, Switzerland}
\affiliation{Quantum Center, ETH Zurich,~CH-8093 Zurich, Switzerland}
\author{Klaus Ensslin}
\affiliation{Solid State Physics Laboratory, ETH Zurich,~CH-8093~Zurich, Switzerland}
\affiliation{Quantum Center, ETH Zurich,~CH-8093 Zurich, Switzerland}

\begin{abstract}
Graphene quantum dots are promising candidates for qubits due to weak spin-orbit and hyperfine interactions. The hyperfine interaction, controllable via isotopic purification, could be the key to further improving the coherence. Here, we use isotopically enriched graphite crystals of both $^{12}$C and $^{13}$C grown by high-pressure-high-temperature method to exfoliate graphene layers. We fabricated Hall bar devices and performed quantum transport measurements, revealing mobilities exceeding $10^{5}$$\textrm{cm}^{2}/Vs$ and a long mean free path of microns, which are as high as natural graphene. Shubnikov-de Haas oscillations, quantum Hall effect up to the filling factor of one, and Brown-Zak oscillations due to the alignment of hBN and graphene are observed thanks to the high mobility. These results constitute a material platform for physics and engineering of isotopically-enriched graphene qubits.
\end{abstract}

\maketitle
\newpage
%%%%%%%%%%%%%%%%%%%%%%%%%%%%%%%%%%%%%%%%%%%%%%%%%%%%%%%%%%%%%
%\section{\label{sec:level1}Introduction}
\section{Introduction}
Graphene quantum dots are among the most promising candidates as platforms for spin qubits\cite{trauzettel_spin_2007, doi:10.1021/acs.nanolett.8b01859,graphene_spin_relaxation} thanks to the weak spin-orbit coupling and hyperfine interactions. Understanding the hyperfine interaction is a strategy to improve the qubit coherence in several systems such as GaAs\cite{petta_coherent_2005,camenzind_hyperfine-phonon_2018,fujita_signatures_2016,erlingsson_hyperfine-mediated_2002} and Si-based systems \cite{eng_isotopically_2015,tyryshkin_electron_2012,Itoh_material}.
The hyperfine interaction has also been investigated in $^{13}$C-enriched carbon nanotube quantum dots \cite{churchill_electronnuclear_2009} and it was theoretically proposed that graphene quantum dots could benefit from isototpe purificiaction.\cite{fischer_hyperfine_2009,sengupta_hyperfine_2023,trauzettel_spin_2007,yazyev_hyperfine_2008}, modulating the ratio of $^{12}$C (nuclear spin 0) and $^{13}$C (nuclear spin 1/2) (see Appendix \ref{section:Coherence time}). Comparing the results of $^{12}$C and $^{13}$C-enrichment would enable the direct investigation of the effect of hyperfine interaction.
While isotopically-enriched graphene has been realized, they have been made by the chemical vapor deposition (CVD) method \cite{strenzke_nuclear-induced_2022,wojtaszek_absence_2014}, whose electronic quality is usually not as high as for exfoliated samples and thus challenging for qubit fabrication.

In this work, we investigate high-quality exfoliated graphene devices out of $^{12}$C-enriched ($99.7 \%$) and $^{13}$C-enriched ($91.4 \%$) graphites. We synthesize enriched graphite by the high-pressure-high-temperature method \cite{taniguchi_spontaneous_2001,taniguchi_synthesis_2007}.
We then establish a recipe to exfoliate the layered crystals to obtain monolayer and bilayer graphene. The isotope effect is characterized by mass spectroscopy and Raman spectroscopy, showing a clear difference between $^{12}$C and $^{13}$C graphene. After confirming the enrichment, we fabricate a Hall bar device from each kind of graphite and perform quantum transport measurements. For both graphene samples, high mobility ($\mu\geq1\times 10^{5}$ cm$^2$/Vs) and long mean free path ($l_\textrm{mfp}\geq 1$ $\mu$m) are demonstrated. These values demonstrate that the electronic quality of graphene remains as high as natural graphene after isotopic enrichment. We also observe the Shubnikov-de Haas oscillations, quantum Hall effect, and Brown-Zak oscillations, which confirm the quality of the sample. This result forms the basis for building isotopically-enriched graphene qubits and investigating the role of hyperfine interactions.

%%%%%%%%%%%%%%%%%%%%%
\begin{figure*}[t]
\centering
\includegraphics{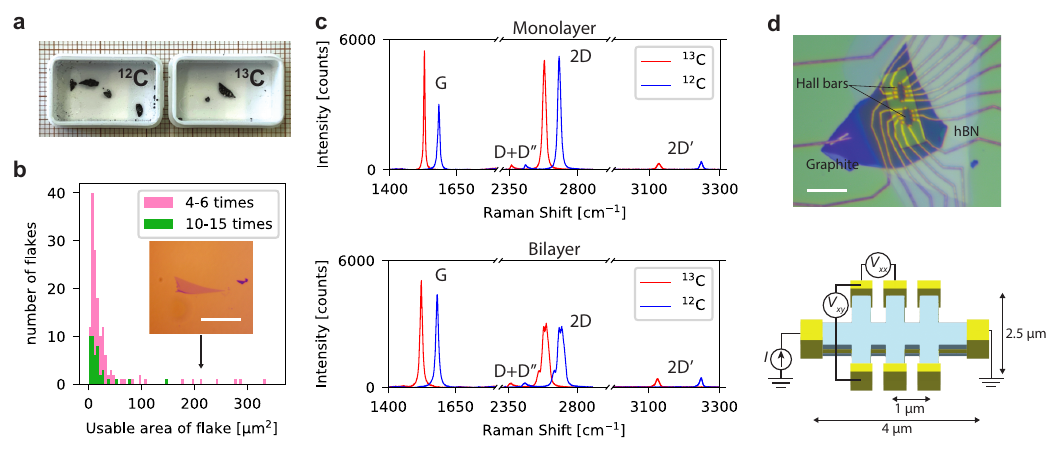}
\caption{(a) Optical picture of $^{12}$C and $^{13}$C graphite crystals. The minimum grid size in the background is 1 mm. (b) Histogram of $^{12}$C graphene flake size. Inset shows a picture of the example bilayer flake. The scale bar is 50 $\mu$m. (c) Raman spectrum of $^{12}$C and $^{13}$C graphene. Top: Monolayer graphene. Bottom: Bilayer graphene. (d) Top panel: Optical picture of the fabricated Hall bar. The scale bar is 10 um. Bottom panel: Schematic of the fabricated Hall bar. Source, drain, and voltage terminals are shown together with the geometric measurements.
}
\label{samplecharac}
\end{figure*}
%%%%%%%%%%%%%%%%%%%%%%%%%%%%%%%%

%%%%%%%introduction%%%%%%%%%%%%%
\section{Characterization Results}
\subsection{Enriched graphite and exfoliation}
Figure \ref{samplecharac}(a) shows the optical picture of the crystal of $^{12}$C-enriched and $^{13}$C-enriched graphite. Hereafter, we call them $^{12}$C graphite/graphene and $^{13}$C graphite/graphene for convenience.
Isotopically enriched graphite crystals are obtained using Co-Ti solvents by a temperature difference method under high pressure. This method modifies the high-pressure synthesis conditions of diamond single crystals in the graphite stability region of 3 GPa and 1600 $^{\circ}$C for 20 hrs \cite{miyakawa_nitrogen_2022}. The carbon source material is dissolved in the molten solvent under high pressure and high temperature and precipitated as a graphite crystal. The source materials are $^{12}$C-enriched diamond crystals prepared by CVD using $^{12}$C-enriched methane gas (Tomei-diamond Co Ltd) and commercially available $^{13}$C-enriched graphite. After high-pressure synthesis, the metallic solvents are removed with hot aqua regia, and the graphite crystals are purified with pure water.

Note that the crystal size ($\sim$5 mm) is relatively small compared to the natural graphite crystals, which can be up to of the order of cm. The relative isotope ratio found in the crystals is determined using inductively coupled plasma time-of-flight mass spectrometry (ICP-TOFMS). The ICP-TOFMS (icp-TOF2R, TOFWERK AG, Thun, Switzerland) is operated in low-mass mode and was coupled to an ArF excimer laser (193 nm, GeoLas C, Lambda Physik, Göttingen), equipped with a low dispersion ablation cell for minimally invasive sampling of the small crystals. The ratios of isotopes are determined for natural graphite, $^{12}$C graphite, and $^{13}$C graphite. The results of the natural graphite are used to calibrate the ratio for the other samples, assuming the natural abundance of $^{12}$C:$^{13}$C = 98.9 : 1.1. The result of the mass spectroscopy yields $^{12}$C:$^{13}$C = 99.7 : 0.3 for the $^{12}$C sample and 8.6 : 91.4 for the $^{13}$C sample, confirming the effect of enrichment. The relative concentrations of the isotopes determined by mass spectroscopy have a relatively large error because of the small amount of the sample available.

In spite of the relatively small size of the graphite crystals, exfoliation is possible for both $^{12}$C and $^{13}$C. We perform exfoliation by using the standard scotch tape method, where we deposit graphite onto the scotch tape and fold and peel apart the tape a certain number of times before the flakes are transferred onto a Si/SiO2 substrate. We compare exfoliation for folding and peeling apart the tape 4-6 times with 10-15 times. We analyze the chips under a microscope and record the size of the flakes. The statistics of the flake size (mono-, bi-, and trilayer $^{12}$C graphene) is shown in Figure \ref{samplecharac}(b), together with an image of an example of a large bilayer flake. It turns out that after 4-6 times of exfoliation, it is possible to obtain 50 $\mu$m scale graphene flakes of mono, bi, and tri layers. The same result is obtained for the $^{13}$C graphene. These results demonstrate that the enriched $^{12}$C and $^{13}$C exfoliation is possible and that the flakes are available for fabricating devices.
%%%%%%%%%%%%%%%%%%%%%%%%%%%%%%%%%%%%%%%%%%%%%%%%%%%
\begin{figure*}[t]
\begin{center}
\includegraphics{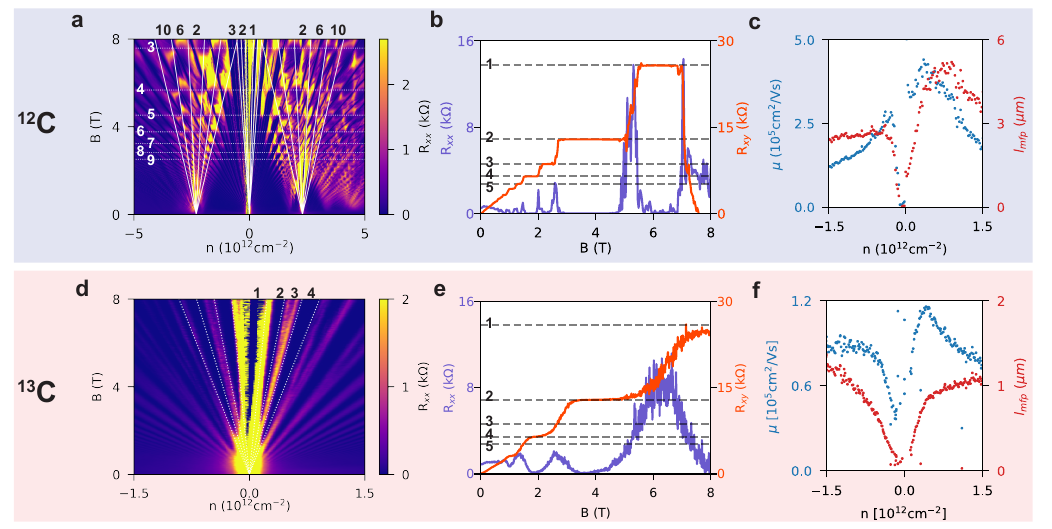}
\caption{Quantum transport measurement of $^{12}$C (a,b,c) and $^{13}$C (d,e,f) bilayer graphene. (a,d) $R_{xx}$ as a function of magnetic field $B$ and carrier density $n$. White dotted tilted lines fit the Shubnikov-de Haas oscillations with filling factors shown in the Figure. White dotted horizontal lines show the position at which the Brown-Zak oscillations appear. (b,e) Example traces of the quantum Hall effect at $n=1.13\times10^{11}$ ($^{12}$C) and $n=1.9\times10^{11}$ ($^{13}$C). (c,f) Carrier density $n$ dependence of the mobility $\mu$ and the mean free path $l_\textrm{mfp}$ estimated from the low magnetic field data.}
\label{LF}
\end{center}
\end{figure*}
%%%%%%%%%%%%%%%%%%%%%%%%%%%%%%%%%%%%%%%%%%%%%%
\subsection{Raman spectroscopy}
The exfoliated graphene is further characterized by Raman spectroscopy (Horiba LabRAM HR Evolution UV-VIS-NIR). The laser energy and wavelength are \SI{1}{mW} and \SI{532}{nm}, respectively. We choose a flake larger than \SI{10}{\micro m} and perform a Raman spectrum measurement by changing the laser position. The data shown in Fig \ref{samplecharac}(c) is from the part of the graphene that is a few \SI{}{\micro m} inside from the edge of the flake. The measurements are done on mono and bilayer flakes.
As shown in Fig. \ref{samplecharac}(c), both $^{12}$C and $^{13}$C graphene show prominent G ($\sim$1520-1580 cm$^{-1}$), D+D$^{''}$ ($\sim$2360-2490 cm$^{-1}$), 2D ($\sim$2540-2710 cm$^{-1}$), and 2D$^{'}$ ($\sim$3120-3250 cm$^{-1}$) peaks. The numbers in the parentheses are typical ranges of the observed peak wavenumbers. We see a clear peak wavenumber shift between $^{12}$C and $^{13}$C. These differences are one of the most prominent signatures of the isotope effect because the nuclear mass difference results in a phonon frequency and Raman shift difference. We do not observe a shift between $^{12}$C and natural graphene within the experimental resolution. After subtracting a linear background from the data, we fit all the peaks with Lorentzian functions. The fit parameters are in the table in Appendix \ref{section:Raman data}.

The obtained parameters are consistent with the ones reported in exfoliated natural graphene and isotopically enriched CVD graphene \cite{SpatialRaman, strainVaration, Bernal_stack, chargeCarrier, TBLGC1213}. The measured wavenumber of the G peak, $\omega_G$, for \textsuperscript{12}C or natural graphene reported in Refs. \cite{SpatialRaman, strainVaration, Bernal_stack, chargeCarrier, TBLGC1213} are between \SI{-2}{cm^{-1}} and \SI{+10}{cm^{-1}} of our measured values for $^{12}$C graphene (for mono- and bilayer). The $\omega_G$ for \textsuperscript{13}C graphene reported in \cite{Bernal_stack, TBLGC1213} is in the range of \SI{-5}{cm^{-1}} to \SI{+8}{cm^{-1}} of the values we measure for mono- or bilayer.
Comparing the wavenumber of the 2D peak, $\omega_{2D}$, to literature values, it turns out that the values reported are generally slightly higher (up to 1.5 $\%$) than the values we measure. For the monolayer \textsuperscript{12}C graphene $\omega_{2D}$, we find that our measurement aligns very well (within \SI{1}{cm^{-1}}) with the natural graphene measurement in \cite{SpatialRaman}. However, the measurements in Refs. \cite{Bernal_stack, chargeCarrier, TBLGC1213} find values 15-\SI{35}{cm^{-1}} higher than what we measure for \textsuperscript{12}C monolayer graphene as well as for \textsuperscript{13}C monolayer graphene. Looking at the bilayer measurements for $\omega_{2D}$ peaks \cite{SpatialRaman}, we find two peaks that are \SI{2}{cm^{-1}} and \SI{3}{cm^{-1}} apart. For $\omega_{2D'}$ \cite{2D'Peaks} reports \SI{3250}{cm^{-1}} for \textsuperscript{12}C graphene and \SI{3130}{cm^{-1}} for \textsuperscript{13}C graphene. These values are consistent with what we measure.

We also estimate the isotopic concentration of $^{13}$C graphene from the Raman shift using the relation $\omega=\omega_\textrm{12C}\sqrt{\frac{m}{m+x\Delta m}}$ \cite{carvalho_probing_2015}. Here, $\omega_\textrm{12C}$ is the Raman shift of pure $^{12}$C graphene, $m$ is atomic mass of $^{12}$C, $x$ is the concentration of $^{13}$C, and $\Delta m$ the atomic mass difference of $^{12}$C and $^{13}$C. 
For $^{13}$C graphene, we obtain a $^{13}$C enrichment $x = 92.9 \%$ using the 2D and 2D' peaks and $x = 86.2 \%$ using the G peak. Since the G peak is sensitive to the distortion and carrier density \cite{das_monitoring_2008}, it can be affected by unintentional doping due to the charged impurity in the silicon substrate, making the peak shift.
The estimation from 2D and 2D' peaks is closer to the value obtained by the mass spectroscopy with a slight overestimation ($\pm 1.5 \%$). At around 93$\%$ of $^{13}$C, the deviation of Raman shifts by one cm$^{-1}$ modulates the concentration estimation by $\sim1.7\%$. Therefore, considering the resolution of the measurement, the estimation from the 2D and 2D' peaks is consistent with the one from the mass spectroscopy.

\subsection{Device fabrication}
To characterize the transport quality of the isotopically enriched graphene, we fabricate Hall bars of $^{12}$C and $^{13}$C bilayer graphene as shown in Figure \ref{samplecharac}(d). A stack of hexagonal boron-nitride (top hBN)/bilayer graphene/hBN (bottom hBN)/graphite (back gate) is made by the poly-dimethylsiloxane/poly-carbonates dry transfer method. Electric contacts to the edge of the bilayer graphene are fabricated by etching the top hBN and depositing Cr/Au. After etching, a Hall bar is shaped by reactive ion etching (CHF$_3$ and O$_2$). We measure the Hall bars in a dilution refrigerator with a base temperature of 55 mK. 
The bottom panel of Fig. \ref{samplecharac}(d) shows the schematic of the device. The two contacts at each sample end are used as source and drain electrodes, where we inject current from the source to the grounded drain. The longitudinal and transverse voltages ($V_{x}$ and $V_{y}$) are measured between the two contacts along and across the sides. The spacing between source and drain, and between the voltage contacts is $\SI{4}{\micro m}$ and $\SI{1}{\micro m}$, respectively.
We use a lock-in amplifier (Stanford Research Systems SR830) connected to the source in series with a \qty{100}{\mega\ohm} resistance and apply an AC voltage of \qty{1}{\volt}, generating an AC current of 10 nA. We synchronize the lock-in amplifier driving the current $I$ with the other two lock-in amplifiers. These lock-in amplifiers are then used to measure the resistances ($R_{xx}=\frac{dV_{x}}{dI}$ and $R_{xy}=\frac{dV_{y}}{dI}$ at zero bias current). We apply a DC voltage (Yokogawa 7651 Programmable DC Source) to the back gate (not shown in the schematic).

\section{Transport measurement}
Figures \ref{LF}(a) and (d) show the magnetic field $B$ and the carrier density $n$ dependence of the longitudinal resistance $R_{xx}$ of $^{12}$C and $^{13}$C devices. 
The carrier density on the horizontal axis is estimated by measuring the classical Hall effect up to 100 mT and using the relation $R_{xy}=\frac{B}{n|e|}$. We also obtain consistent values from a parallel plate capacitor model ($n=C_\textrm{bg}V_\textrm{bg}$, where $C_\textrm{bg}$ is the capacitance between the graphene and the back gate and $V_\textrm{bg}$ is the back gate voltage) and Shubnikov-de Haas measurements ($R_{xx}\propto \textrm{cos}(\frac{2\pi nh}{4eB})$).

For both samples, clear fan-like structures departing from the charge neutrality point $n=0$ and expanding with magnetic field are seen, which is attributed to the Shubnikov-de Haas (SdH) oscillations. The oscillations appear already at around 1 T, testifying to our samples' high mobility. White dotted tilted lines show the fitting to the filling factor of 1, 2, and 3 at around $n=0$ (for $^{12}$C and $^{13}$C) and filling factor for 2, 6, and 10 at around $n=\pm2.3\times10^{12}$ cm$^{-2}$ (for $^{12}$C only).

Additionally, in the $^{12}$C sample, we see multiple Landau fans appearing at densities of around 4.09, 3.29, 2.30, 1.78, -2.06, and -2.31 $\times10^{12}$cm$^{-2}$. This is attributed to the additional Dirac points due to the unintended alignment between the graphene and one of the hBNs, forming a moiré superlattice \cite{Dirac_points_superlattice}. 
%\qtylist{-6.4; -5.1; -3.6; -2.9; 2.2; 2.8}{\volt}
This results in an energy spectrum for the charge carriers known as the Hofstadter butterfly and causes satellite Dirac peaks. Furthermore, at the intersections of the Landau fans, it predicts horizontal lines of peaks in $R_{xx}$ called Brown-Zak oscillations \cite{exp_Hofstadter_butterfly,Brown_Zak}. The horizontal lines in Fig. \ref{LF}(a) indicate the position of $\phi/\phi_0=1/p$ with $p$ an integer and $\phi_0=h/e$ the flux quanta. As we discuss in Appendix~\ref{BZ}, the moir\'e unit cell size estimated from the Brown-Zak oscillations agrees with the one formed by graphene/hBN alignment. This observation is another piece of evidence for having a high-quality sample.

At a high magnetic field, we observe the quantum Hall effect, as shown in Figs.~\ref{LF}(b) and (e). For both samples, the integer quantum Hall effect is observed up to filling factor $\nu=1$ with quantized plateaus of $R_{xy}$ and dips of $R_{xx}$.
%When $B\leq 4$ T, the filling factor changes in a step of four. As the magnetic field is increased, the spacing becomes one. This transition suggests that the four-fold degeneracy of spin and valley is lifted due to the high magnetic field and low temperature. 
The significant drop of $R_{xy}$ in $^{12}$C beyond filling factor 1 ($B\geq7$ T) is due to the Brown-Zak oscillations. We observe a similar drop whenever $B$ crosses the horizontal lines indicated in Fig. \ref{LF}(a).

We also estimate the mobility $\mu$ and mean free path $l_\textrm{mfp}$ by applying the classical Drude model to $\rho_{xx}=\frac{W}{L}R_{xx}$ ($W$ and $L$ are the width and the length of the sample) and $\rho_{xy}=R_{xy}$ at a low magnetic field ($\leq 100$ mT). From the slope of the linear fit of $R_{xy}$ up to 100 mT together with $R_{xx}$ at \qty{0}{\tesla}, we determine $\mu$ and $l_\textrm{mfp}$ using the equations $\mu=\frac{1}{\rho_{xx}(B=0)}\frac{d\rho_{xy}}{dB}$ and $l_\textrm{mfp}=\frac{\hbar\sqrt{\pi n}}{|e|}\mu$. As seen in Figure \ref{LF}(c) and (f), the mobility mostly ranges from $1\times10^{4}$ to $\SI{3e5}{cm^2/Vs}$. The mobility becomes zero around charge neutrality and forms a peak with increasing $n$. This behavior can be attributed to the difference in dominant scattering mechanisms (long-range Coulomb scattering at low density and short-range impurity scattering at high density) \cite{SLG_mobility_model}.

The mean free path $l_\textrm{mfp}$ reaches up to $1-\SI{5}{\micro m}$ for both $^{12}$C and $^{13}$C. This value is comparable to the one reported in natural graphene. Note that the spacing between the contacts is of the same order of magnitude as the estimated mean free path, meaning that the transport in the sample is in the ballistic regime. In this regime, the Drude model has limited validity, and the mobility and mean free path are lower bound estimates. 

We further perform a quantum Hall effect breakdown measurement, applying a DC current up to \SI{10}{\micro A} so that the quantum Hall effect (plateau in $\rho_{xy}$ and zero in $\rho_{xx}$) is no longer observed. In GaAs 2DEG systems, a large hysteresis in the current sweep direction before and after the breakdown is observed typically for odd filling factors due to dynamic nuclear polarization (spin transfer from electrons to nuclei) \cite{song_strong_2000,kawamura_electrical_2007}. We do not observe any clear hysteresis within our measurement precision (data is not shown here). To observe the effect of hyperfine interaction, more sophisticated measurements such as resistivity-detected NMR \cite{yusa_controlled_2005,stern_nmr_2004,desrat_resistively_2002} or actually building a quantum dot and performing a $T_1$ and $T_2$ coherence time measurement would be useful.

%%%
\section{Conclusion}
In conclusion, we have presented high-mobility transport of exfoliated $^{12}$C and $^{13}$C graphene synthesized by the high-pressure-high-temperature technique. We identified distinct differences between $^{12}$C and $^{13}$C graphene by the mass and Raman spectroscopy. We also fabricated Hall bar devices and performed quantum transport measurements, revealing high mobility and a long mean free path. Shubnikov-de Haas, quantum Hall, and Brown-Zak effects up to the filling factor of one were observed thanks to the high mobility. These results pave the way for developing isotopically-enriched graphene qubits and investigating the role of hyperfine interactions in graphene quantum dots.

\begin{acknowledgements}
We are grateful for fruitful discussions and technical support from Lev Ginzburg, Chuyao Tong, Markus Niese, Peter Maerki, Thomas Baehler, and the ETH FIRST cleanroom facility staff. We acknowledge financial support by the European Graphene Flagship Core3 Project, H2020 European Research Council (ERC) Syn- ergy Grant under Grant Agreement 951541, the European Union’s Horizon 2020 research and innovation programme under grant agreement number 862660/QUANTUM E LEAPS, the European Innovation Council under grant agreement number 101046231/FantastiCOF, NCCR QSIT (Swiss National Science Foundation, grant number 51NF40-185902).
K.W. and T.T. acknowledge support from the JSPS KAKENHI (Grant Numbers 21H05233 and 23H02052) and World Premier International Research Center Initiative (WPI), MEXT, Japan.
\end{acknowledgements}

%%%%%%%%%%%%%%%%%%%%%%%%%%%%%%%%%%%%%%%%%%%%%%%%%%%%%%%%%%%%%%%%%%%%%%%%%%%%%%%%%%%%%%%%%%%%%%%%%%%%%%%%%%%%%%%%%%%%%%%%%%%%%%%%%%%%%%%%%%%%%%%%%%%%%%%%%%%%%%%%%%%%%%%%%%%%%%%%%%%%%%%%%%%%%%%%%%%%%%%%%%%%%%%%%%%%%%%%%%%%%%%%%%%%%%%%%%%%%%%%%%%%%%%%%%%%%%%%%%%%%%%%%%%%%%%%%%%%%%%%%%%%%%%%%%%%%%%%%%%%%%%%%%%%%%%%%%%%%%%%%%%%%%%%%%%%%%%%%%%%%%%%%%%%%%%%%%%%%%%%%%%%%%%%%%%%%%%%%%%%%%%%%%%%%%%%%%%%%%%%%%%%%%%%%%%%%%%%%%%%
\setcounter{equation}{0}
\setcounter{figure}{0}
\renewcommand{\theequation}{A.\arabic{equation}}
\renewcommand{\thefigure}{A.\arabic{figure}}

\revappendix
\section{Coherence time estimation}
\label{section:Coherence time}
Following the steps of \cite{trauzettel_spin_2007}, we studied the effect of isotopic composition on the lifetime $\tau$ of coherent spin states ($T_\textrm{2}$) in graphene quantum dot qubits. While the different atomic masses of ${^{12}}$C and ${^{13}}$C may affect the spin-orbit coupling, it is believed that the hyperfine interaction limits the coherence time in graphene \cite{trauzettel_spin_2007}. The limit on the coherence time set by the hyperfine interaction (hyperfine coherence time) $\tau_{\text{hf}}$ depends on the composition of the spinless ${^{12}}$C and the spin-1/2 ${^{13}}$C. It can be estimated from the hyperfine coupling strength $A_{\text{hf}} = \qty{0.38}{\micro \electronvolt}$, the ${^{13}}$C concentration $c_{13C}$ and the number of atom in a dot $N$ using $\tau_{\text{hf}} = \frac{h}{A_{\text{hf}}}\sqrt{{N}/{c_{13C}}}.$

Compared to natural graphene ($c_{13C} = \qty{1.1}{\percent}; N = \qty{e4}{} : \tau_{\text{hf}} \approx \qty{10}{\micro \second}$) the purification level of our ${^{12}}$C sample ($c_{13C} = 0.3 \%$) would already increase the hyperfine coherence time by a factor of two. Further decreasing the ${^{13}}$C concentration (99.99\% purified ${^{12}}$C), would increase the coherence time by another factor of five, to around $\tau_{\text{hf}} \approx \qty{100}{\micro \second}$. This shows the strong sensitivity of the hyperfine coherence time on isotopic composition at low ${^{12}}$C concentrations. For a dot size of $N = \qty{e4}{}$ atoms, this purification level of 99.99\% would already mean that on average just one ${^{13}}$C atom would be present per dot. By selecting the dots without any ${^{13}}$C and thus no nuclear spins, one can completely lift the limitation of the coherence time by the hyperfine interaction. 
%%%%%%%%%%%%%%%%%%%%%%%%%%%%%%%%%%%%%%%%%%%%%%%%%%%%%%%%%%%%%%%%%%%%%%%%%%%%%%%%%%%%%%%%%%%%%%%%%%%%%%%%%%%%%%%%%%%%%%%%%%%%%%%%%%%%%%%%%%%%%%%%%%%%%%%%%%%%%%%%%%%%%%%%%%%%%%%%%%%%%%%%%%%%%%%%%%%%%%%%%%%%%%%%%%
%%%%%%%%%%%%%%%%%%%%%%%%%%%%%%%%%%%%%%%%%%%%%%%%%%%%%%%%%%%%%%%%%%%%%%%%%%%%%%%%%%%%%%%%%%%%%%%%%%%%%%%%%%%%%%%%%%%%%%%%%%%%%%%%%%%%%%%%%%%%%%%%%%%%%%%%%%%%%%%%%%%%%%%%%%%%%%%%%%%%%%%%%%%%%%%%%%%%%%%%%%%%%%%%%

\section{Raman peaks fitting parameters}
\label{section:Raman data}

We fitted the Raman spectrum with Lorenzian function $\frac{A\Gamma^2}{(\omega-\omega_0)^2+\Gamma^2}$ where $A$ is the peak amplitude, $k$ is the wavenumber, $k_0$ is the peak wavenumber, and $\Gamma$ is the linewidth (half-width-half-maximum). The results are shown below. We used a single Lorenzian function for $G$ and $2D^{'}$ peaks, sum of two Lorenzian functions for $D+D^{''}$ peak, and sum of three Lorenzian functions for $2D$ peak.

\begin{table}[h]
    \centering
    \begin{tabular}{| c | m{1cm} | m{1cm} | m{1cm} | m{1cm} |}
            \hline 
         Layers & 1 & 1 & 2 & 2 \\ \hline
         Material & \textsuperscript{12}C & \textsuperscript{13}C & \textsuperscript{12}C & \textsuperscript{13}C \\ \hline
         $\omega_G [\textrm{cm}^{-1}]$ &  1584 & 1530	& 1581 & 1522 \\ \hline
         $\Gamma_G [\textrm{cm}^{-1}]$ &  8.143 & 5.316 & 11.14 & 9.983	\\ \hline
         $\omega_{2D} [\textrm{cm}^{-1}]$ &  2678 & 2580 & 2646, 2681, 2704 & 2546, 2580, 2602 \\ \hline
         $\Gamma_{2D} [\textrm{cm}^{-1}]$ &  21.431 & 22.296 & 14.98, 21.76, 31.32 & 13.49, 21.47, 29.96 \\ \hline
         $\omega_{2D'} [\textrm{cm}^{-1}]$ &  3247 & 3128	& 3248	& 3125	\\ \hline
         $\Gamma_{2D'} [\textrm{cm}^{-1}]$ &	8.455 & 11.274 & 8.709 & 10.61 \\ \hline
         $\omega_{D+D''} [\textrm{cm}^{-1}]$ & 2479, 2457&  2366, 2388& 2458, 2489& 2365, 2394\\ \hline
         $\Gamma_{D+D''} [\textrm{cm}^{-1}]$ & 41.27, 15.29 & 14.41, 36.46 & 26.04, 48.69 & 21.82, 41.55\\ \hline
         \end{tabular}
        \caption{Table of fit parameters for the Raman peaks. We fit the data with the sum of one to three Lorentzian functions (in the case we use multiple Lorenzian to fit, all fit parameters are shown in order and separated by comma).}
        \label{tab: raman peak param}
\end{table}

\section{Estimation of moire unit cell size from the Brown-Zak oscillation}
\label{BZ}
We have estimated the moire cell size with two approaches. 
First, since the satellite Dirac peaks appear once all the available states in the moiré cell are filled, we can determine the area $A$ of the moiré cell from the average of the excess carrier density of the electron- and hole-side peaks $n_{ep} = \qty{1.78e12}{\per \cm \squared}$ and $n_{hp} = \qty{-2.31e12}{\per \cm \squared}$.
Considering the 4-fold degeneracy in graphene, this condition corresponds to $A = \frac{4}{\frac{1}{2}(|n_{ep}| + |n_{hp}|)}$. 
The second approach uses the periodicity of the Brown-Zak oscillations, which appear if a rational number $1/p$ of flux quanta $\phi_0$ (p is an integer) passes through one moiré cell. The moiré cell area can then be determined from the positions of two adjacent Brown-Zak peaks in magnetic field ($B_1$ and $B_2$) using $A = \phi_0 \left|\frac{1}{B_1} - \frac{1}{B_2} \right|$. The lattice constant can be obtained from the area by simple geometric considerations as $L = \sqrt{2A/\sqrt{3}}$. Both methods give a consistent estimate of the moire unit cell size of 10 nm, which is reasonable for the graphene/hBN-aligned system. 

\bibliography{main_bib}

\end{document}